\documentclass{webofc}
\usepackage{amssymb}
\usepackage{hyperref}
\usepackage[varg]{txfonts}   % Web of Conferences font
\usepackage{url}
\hypersetup{colorlinks=true,citecolor=blue,urlcolor=blue,linkcolor=blue,pdftitle={The {(3+1)D} structure of the dilute Glasma},pdfauthor={Andreas Ipp, Markus Leuthner, David I. M\"uller, Pragya Singh, S\"oren Schlichting, Kayran Schmidt}}
\begin{document}
\title{The (3+1)D structure of the dilute Glasma}
%
% subtitle is optional

\author{%
\firstname{Andreas} \lastname{Ipp}\inst{1}\fnsep\thanks{\email{ipp@hep.itp.tuwien.ac.at}}%
\and%
\firstname{Markus} \lastname{Leuthner}\inst{1}%
\and%
\firstname{David I.} \lastname{M\"uller}\inst{1}\fnsep\thanks{\email{dmueller@hep.itp.tuwien.ac.at}}%
\and%
\firstname{S\"oren} \lastname{Schlichting}\inst{2}\fnsep\thanks{\email{sschlichting@physik.uni-bielefeld.de}}%
\and%
\firstname{Kayran} \lastname{Schmidt}\inst{1}\fnsep\thanks{Speaker, \email{kschmidt@hep.itp.tuwien.ac.at}}%
\and%
\firstname{Pragya} \lastname{Singh}\inst{3,4}
}

\institute{%
Institute for Theoretical Physics, TU Wien, Wiedner Hauptstraße  8-10/136, A-1040 Vienna, Austria
\and
Fakult\"at f\"ur Physik, Universit\"at Bielefeld, D-33615 Bielefeld, Germany
\and
Department of Physics, University of Jyv\"askyl\"a, FI-40014 Jyv\"askyl\"a, Finland
\and
Helsinki Institute of Physics, University of Helsinki, FI-00014 Helsinki, Finland
}

\abstract{
We study the (3+1)D structure of the Glasma in the dilute approximation, which allows us to describe the longitudinal dynamics that arise from the three-dimensional nuclear structure.
We employ a nuclear model with tunable longitudinal and transverse fluctuation scales that generalizes the McLerran-Venugopalan model.
We discuss the longitudinal profiles of the energy-momentum tensor and the transverse structure of the local rest frame energy density.
}
\maketitle

\section{Introduction}\label{sec:intro}

Tremendous experimental and theoretical effort has led to a \emph{standard} picture~\cite{ALICE:2022wpn} of relativistic heavy-ion collisions (HICs).
In the initial state, the nuclei are described by the color glass condensate (CGC)~\cite{Gelis:2010nm,Gelis:2012ri} that is based on a separation of scales into hard valence partons and soft gluon fields.
After the collision, the CGC evolves according to classical Yang-Mills (YM) equations into the Glasma~\cite{Lappi:2006fp}, which thermalizes into a quark-gluon plasma that eventually hadronizes into particles.
The initial state and Glasma are traditionally described in the boost-invariant limit, which leads to the loss of longitudinal dynamics.
However, the importance of the longitudinal structure in the initial state and Glasma drives ongoing research.

In these proceedings, we discuss our three-dimensional nuclear model with tunable correlation scales that goes beyond the boost-invariant description of the Glasma.
We summarize the dilute (3+1)D Glasma framework, which allows us to calculate the field strength tensor of the Glasma while keeping the longitudinal dependence fully general, and discuss results using our nuclear model.

%----------
\section{The dilute Glasma}\label{sec:glasma}

The CGC approach to model the initial state of the highly-relativistic nuclei is formulated in terms of the classical color currents $\mathcal{J}^\mu_{A/B}(x^\pm,\mathbf{x}) = \delta^\mu_\mp\,\rho_{A/B}(x^\pm,\mathbf{x})$, which are assumed to be recoilless and frozen in the light cone times $x^\mp=(t\mp z)/\sqrt{2}$ of the left-moving nucleus $A$ (upper signs) and right-moving $B$ (lower signs).
The color charge distributions $\rho_{A/B}(x^\pm,\mathbf{x})$ of the hard partons in the nuclei vary in the transverse plane $\mathbf{x}=(x,y)$ and have finite, non-trivial support along the longitudinal coordinates $x^\pm$.
These currents source the soft gluon fields $\mathcal{A}_{A/B}^\mu(x^\pm,\mathbf{x})$ in the nuclei, which are given as solutions of the classical YM equations in covariant gauge and enter the single-nucleus field strengths $\mathcal{F}^{i\mp}_{A/B}(x^\pm,\mathbf{x}) = \partial^i \mathcal{A}^\mp_{A/B}(x^\pm,\mathbf{x})$.

The description of the collision problem of two such single-nucleus currents and fields requires solutions of the classical YM equations in terms of the field strength tensor of the gauge field $A^\mu(x)$ that develops in the interaction region of the collision.
The dilute (3+1)D Glasma framework~\cite{Ipp:2021lwz,Ipp:2024ykh,Leuthner:2025vsd} provides the ansatz (in covariant gauge)
$A^\mu(x) = \mathcal{A}^\mu_A(x^+,\mathbf{x}) + \mathcal{A}^\mu_B(x^-,\mathbf{x}) + a^\mu(x)$
and identifies the non-linear contributions from $a^\mu(x)$ as the Glasma field.
Performing an expansion in $\rho_A^m\rho_B^n$ up to leading order $m=n=1$ allows for solutions of the YM equations in terms of an inhomogeneous wave equation for the Glasma field strength tensor $f^{\mu\nu}(x)$.
The known single-nucleus $\mathcal{F}^{i\mp}_{A/B}(x^\pm,\mathbf{x})$ enter the sources of the wave equation.
In this leading order of the expansion, the field strength tensor is effectively abelian $f^{\mu\nu}(x) = \partial^\mu a^\nu(x) - \partial^\nu a^\mu(x)$.
We presented concise expressions for $f^{\mu\nu}(x)$ in~\cite{Ipp:2024ykh} that are cast into integrals of commutators of $\mathcal{F}^{i\mp}_{A/B}$ covering the backward light cone that is attached to the position of an observer at any spacetime point $x$ in the future light cone of the collision.

\section{Nuclear model}\label{sec:nucl}

\begin{figure}
\centering
\sidecaption
\includegraphics[width=0.52\textwidth]{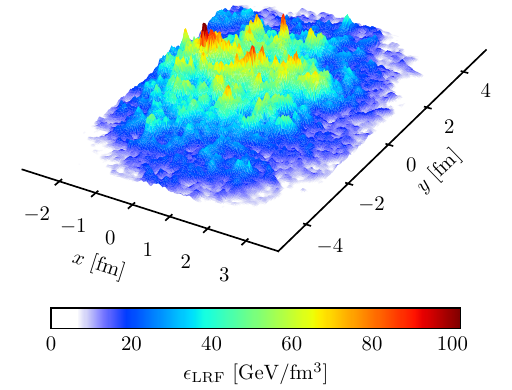}
\caption{Local rest frame energy density $\epsilon_\mathrm{LRF}$ in the transverse $\mathbf{x}=(x,y)$ plane at mid-rapidity $\eta_s=0$ and proper time $\tau=0.4$~fm/c for a single event with impact parameter equal to $R$ and $\sqrt{s_\mathrm{NN}}=200$~GeV corresponding to Au+Au collisions at RHIC.
The IR regulator $m=0.2$~GeV sets the size of the fluctuating domains to $\sim1/m$.}
\label{fig:glasma-relief}
\vspace{-0.5\baselineskip}
\end{figure}

The charge distributions $\rho_{A/B} = \rho^a_{A/B} t^a$, where $t^a$ are the generators of $\mathfrak{su}(3)$, vary stochastically event-by-event.
We use a generalization of the McLerran-Venugopalan (MV)~\cite{McLerran:1993ka,McLerran:1993ni} model that provides realizations of $\rho^a_{A/B}$ according to a Gaussian probability functional fixed by $\langle \rho^a(x^\pm, \mathbf{x}) \rangle = 0$ and
\begin{align}
    \langle \rho^a(x^\pm,\mathbf{x}) \rho^b(y^\pm,\mathbf{y})\rangle &= g^2\mu^2\delta^{ab}\sqrt{T(x^\pm, \mathbf{x})}\sqrt{T(y^\pm, \mathbf{y})} U_\xi(x^\pm-y^\pm)\delta^{(2)}(\mathbf{x}-\mathbf{y}), \label{eq:correlator_factorized}
\end{align}
with the coupling $g$ and MV parameter $\mu$, which has units of energy.
Here, we added three-dimensional, finite, single-nucleus envelopes
\begin{align}
    T(x^\pm, \mathbf{x}) = c \left(1+\exp(\frac{\sqrt{2(\gamma \, x^\pm)^2+\mathbf{x}^2}-R}{d})\right)^{-1}, \qquad c = \frac{\gamma}{\sqrt{2}d\ln(1+\mathrm{e}^{R/d})},
\end{align}
where $R$ and $d$ are the nuclear radius and skin depth and $\gamma=\sqrt{s_\mathrm{NN}}/(2m_u)$ depends on the collider energy and nucleon mass.
We introduce
\begin{align}
    \label{eq:Uxi}
    U_\xi(x^\pm-y^\pm) = \frac{1}{\sqrt{2\pi \xi^2}} \exp(\frac{(x^\pm-y^\pm)^2}{4R^2/\gamma^2}) \exp(-\frac{(x^\pm - y^\pm)^2}{2\xi^2}),
\end{align}
as a factor to tune longitudinal correlations with the scale parameter $\xi\leq \sqrt{2}R/\gamma$.
Given a realization of the nuclear charge density, we solve for $\mathcal{A}^{\mp}$ on a lattice in momentum space
\begin{align}
    \mathcal{A}^{\mp,a}(x^\pm, \mathbf{x}) = \frac{1}{(2\pi)^2} \int d^2 \mathbf{k}\, \frac{\tilde \rho^a (x^\pm, \mathbf{k})}{\mathbf{k}^2+m^2}\mathrm{e}^{-\mathbf{k}^2/(2\Lambda_\mathrm{UV}^2)}\mathrm{e}^{-\mathrm{i}\mathbf{k}\cdot\mathbf{x}},
\end{align}
with the ultraviolet regulator $\Lambda_\mathrm{UV} = 10$~GeV chosen adequately for the lattice spacing and infrared (IR) regulator $m$ as a model parameter.
On the level of the single-nucleus fields $\mathcal{A}^\mp$, $1/m$ controls the size of the fluctuating domains in the transverse plane.

\section{Event-by-event results}\label{sec:results}

We solve the integrals for the field strength tensor $f^{\mu\nu}(x)$ numerically event-by-event with our custom, highly-parallel Monte Carlo integration routines~\cite{Ipp:2021lwz} on a grid for $x=(\tau,\eta_s,\mathbf{x})$, where $\tau=\sqrt{2x^+x^-}$ is the proper time and $\eta_s=\ln(x^+/x^-)/2$ is the spacetime rapidity in the coordinate Milne frame.
We study the energy momentum tensor $T^{\mu\nu}$ and local rest frame energy density $\epsilon_\mathrm{LRF}$ of the Glasma given by the Landau condition
\begin{equation}
    T^{\mu\nu} = 2 \mathrm{Tr}[ f^{\mu\rho}f_\rho^{\hphantom{\rho}\nu} + \tfrac{1}{4}g^{\mu\nu}f^{\rho\sigma}f_{\rho\sigma} ], \qquad T^\mu_{\hphantom{\mu}\nu}\, u^\nu = \epsilon_\mathrm{LRF}\, u^\mu.
\end{equation}

In Fig.~\ref{fig:glasma-relief} we show $\epsilon_\mathrm{LRF}$ for a single Au+Au collision event at RHIC energy.
The $\epsilon_\mathrm{LRF}$ is only deposited within the overlap region of the nuclear profiles in the transverse plane.
In contrast to Glauber, MC-KLN and IP-Glasma results~\cite{Schenke:2012wb}, the size of the transverse domains is governed by $1/m$, as it was the case for the single-nucleus fields.
This is because in the dilute approximation, there does not develop a saturation scale $Q_s \propto g^2 \mu$ as it does for non-perturbative calculations.
Instead, the IR regulator introduced in the nuclear model remains as the only transverse scale in the dilute Glasma and $g^2\mu$ becomes an overall prefactor.

\begin{figure}[t]
\centering
\sidecaption
\hspace*{0.035\textwidth}\includegraphics[width=0.45\textwidth]{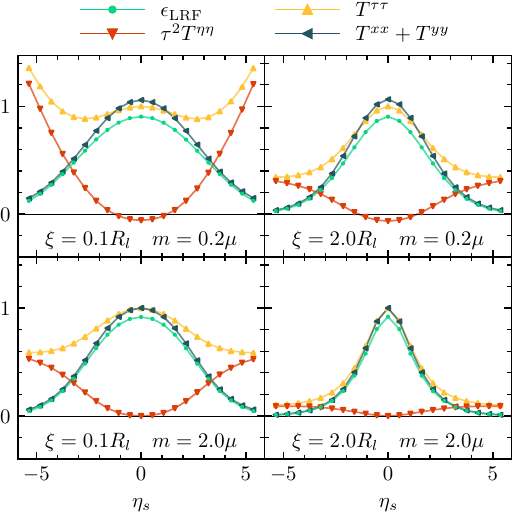}\hspace*{0.035\textwidth}
\caption{%
Rapidity ($\eta_s$) profiles for $\epsilon_\mathrm{LRF}$ and different components of $T^{\mu\nu}$ integrated over the transverse plane and averaged for 10 central collision events at proper time $\tau=0.4$~fm/c corresponding to Au+Au collisions at RHIC.
The Milne frame tensor components of $T^{\mu\nu}$ show unphysical enhancement for large $\eta_s$ and are not suited to describe collisions with extended longitudinal structure.
The sum of transverse pressures $T^{xx}+T^{yy}$ and $\epsilon_\mathrm{LRF}$ are unaffected by the transformation to the Milne frame. Figure from~\cite{Ipp:2024ykh}.%
}
\label{fig:T-profiles}
\vspace{-.5\baselineskip}
\end{figure}

Our results for the transverse integrated rapidity profiles of $T^{\mu\nu}$ and $\epsilon_\mathrm{LRF}$ are shown in Fig.~\ref{fig:T-profiles}.
We identify an enhancement of the $T^{\tau\tau}$ component for large $\eta_s$ compared to $\epsilon_\mathrm{LRF}$ or $T^{xx}+T^{yy}$.
We explain this enhancement as a coordinate effect arising from large longitudinal pressures in the co-moving Bjorken frame.
These pressures lead to growing $T^{\tau\tau}$ because $T^{\mu\nu}$ is traceless in any frame.
We argue that for HICs with extended, non-trivial longitudinal structure in the initial conditions, the co-moving Bjorken frame in Milne coordinates is too impractical to adequately describe the dynamics.

\section{Conclusion}\label{sec:conclusion}
We discussed the dilute Glasma as the leading-order result of the expansion in color charge distributions of the classical YM equations that describe the evolution of gluon fields after the collision of two relativistic heavy ions.
In this treatment, the field strength tensor of the Glasma reduces to concise expressions that allow for non-trivial longitudinal structure in the CGC model for the nuclei.
We used a three-dimensional generalization of the MV model with finite longitudinal extent and tunable transverse and longitudinal correlation scales.
Our results are obtained from event-by-event calculations of the field strength tensor.
We discussed rapidity profiles of $T^{\mu\nu}$ and $\epsilon_\mathrm{LRF}$ and identified the unphysical enhancement of $T^{\tau\tau}$ for large $\eta_s$ as a coordinate effect in the Milne frame.
This suggests that the Milne frame is not suited for calculations using initial conditions with extended longitudinal structure.
Finally, we characterized the structure of $\epsilon_\mathrm{LRF}$ in the transverse plane in terms of the correlation scale $1/m$, which is closely related to a saturation scale in the dilute approximation.

%----------
\vspace{0.5\baselineskip}
\begin{acknowledgement}
\footnotesize
\hyphenation{Deut-sche}
\hyphenation{For-schungs-ge-mein-schaft}
\hyphenation{Dok-to-rats-kol-leg}
ML, DM and KS are supported by the Austrian Science Fund FWF No.\ P34764.
DM is further supported by the Austrian Science Fund FWF No.\ 10.55776/PAT3667424.
SS acknowledges support by the Deutsche Forschungsgemeinschaft (DFG, German Research Foundation) through the CRC-TR 211 \lq Strong-interaction matter under extreme conditions\rq\ – project number 315477589 – TRR 211.
The computational results have been achieved in part using the Austrian Scientific Computing (ASC) infrastructure.
\end{acknowledgement}

%
% BibTeX or Biber users please use (the style is already called in the class, ensure that the "woc.bst" style is in your local directory)
\bibliography{bib} % Replace "your_bib_file" with the actual name of your .bib file

\end{document}